\documentclass[iop]{emulateapj-rtx4} 
\shortauthors{Sekanina}
\shorttitle{`Oumuamua's Peculiar Acceleration}
\slugcomment{Version \today }

\newcommand{\lapeq}{$\;$\raisebox{0.3ex}{$<$}\hspace{-0.28cm}\raisebox{-0.75ex}{$\sim$}$\;$}

\begin{document}
\title{Outgassing As Trigger of 1I/`Oumuamua's Nongravitational Acceleration:\\Could
 This Hypothesis Work At All?}
\author{Zdenek Sekanina}
\affil{Jet Propulsion Laboratory, California Institute of Technology,
  4800 Oak Grove Drive, Pasadena, CA 91109, U.S.A.}
\email{Zdenek.Sekanina@jpl.nasa.gov{\vspace{0.15cm}}}

\begin{abstract}
The question of what triggered the nongravitational acceleration of
1I/`Oumuamua continues~to~attract researchers' attention.  The
absence of any signs of activity notwithstanding, the prevailing notion
is that the acceleration of the stellar, cigar-like object
was prompted by outgassing.~However,~the Spitzer Space Telescope's
failure to detect `Oumuamua not only ruled out the CO$_2$ and/or CO
driven activity (Trilling et al.\ 2018), but made the cigar shape
incompatible with the optical observations.  Choice of water ice as
the source of outgassing is shown to be flawed as well:\ (i)~the water
sublimation law is demonstrably inconsistent with the observed variations
in the nongravitational~acceleration derived by Micheli et al.\ (2018),
{\vspace{-0.04cm}}the point that should have been assertively highlighted;
and (ii)~an upper limit of the production rate of water is estimated at as
low as 4\,$\times$\,10$^{23}$\,molecules s$^{-1}$, requiring that, at most,
only a small area of the surface be active.  In this case the conservation
of momentum law is satisfied only when `Oumuamua's bulk density is extremely
low, $<$0.001~g~cm$^{-3}$,~reminiscent~of the formerly proposed scenario with
`Oumuamua as a fragment of a dwarf interstellar comet, possibly an embryo
planetesimal, disintegrating near
perihelion, with the acceleration driven by solar radiation pressure (Sekanina
2019a) and no need for activity at all. 
High quality of astrometry~and~Micheli~et~al.'s orbital analysis,
whose results were confirmed by the computations of other authors, is acknowledged.
\end{abstract}


\section{Introduction}
Micheli et al.'s (2018) discovery of a nongravitational acceleration
affecting the orbital motion~of~\mbox{`Oumuamua} continues to be attributed
to outgassing, even though all efforts to detect a trace of activity failed.
This situation prevails in part because a 3$\sigma$ upper limit of the
production rate~of water, \mbox{$Q_{{\rm H}_2{\rm O}} < 1.7 \!\times\;\!\!\!
10^{27}$molecules s$^{-1}$}, derived~by~Park et al.\ (2018) from their
nondetection of the OH radio lines near 18~cm on 2017 November~12.97~UT
(with the object 1.80~AU from the Sun), is rather soft.  Ye et al.\ (2017)
presented tighter 3$\sigma$ upper limits on the production rates of three
molecular species with emission bands at the optical wavelengths:\
\mbox{$Q_{\rm CN} \!<\:\!\! 2 \:\!\! \times \! 10^{22}$}, \mbox{$Q_{{\rm C}_2}
\!<\:\!\! 4 \:\!\! \times \! 10^{22}$}, and \mbox{$Q_{{\rm C}_3} \!<\:\!\!
2 \:\!\! \times \! 10^{21}$\,molecules s$^{-1}$}, from their spectroscopic
observations on October 26.21~UT, when the object's heliocentric distance
equaled 1.39~AU.  Very important 3$\sigma$ constraints were reported by
Trilling et al.\ (2018) on the production rates of carbon{\vspace{-0.04cm}}
dioxide, \mbox{$Q_{{\rm CO}_2} \!<\:\!\! 9 \:\!\!\times \! 10^{22}$\,molecules
s$^{-1}$}, {\vspace{-0.04cm}}and of carbon~monoxide, \mbox{$Q_{\rm CO}
\mbox{\lapeq} 9 \:\!\!\times \!  10^{21}$\,molecules s$^{-1}$}, determined
from their non\-detection of `Oumuamua with the Spitzer Space Telescope during
a set of exposures spanning nearly 33~hours and centered on November~22.11~UT,
with the object then 2.00\,$\pm$\,0.015~AU from the Sun.  I show in Section~3
that all these limits can be used to provide independent constraints
on the production of water.

\section{Red Flag Ignored}
%
The comprehensive study by Micheli et al.~(2018)~offers much insight into modeling
the sublimation process and provides some useful constraints on the nature~of~the
relevant mechanism by stating explicitly that:\ (i)~the nongravitational effect
is adequately described by a radial acceleration only, with no need to incorporate
the other two components; (ii)~the introduction of discontinuities in the motion
does not improve the data fit; and (iii)~the variations with heliocentric distance
{\vspace{-0.01cm}}$r$ are consistent with $r^{-2}$ or $r^{-1}$ but not with any
steeper laws, including the Style~II model introduced by Marsden et al.\
(1973) to describe the nongravitational effects generated by the sublimation of
water ice.\footnote{Marsden et al.'s Style II model has successfully been employed
to fit the orbital motions of a large number of comets in both short-period
and nearly-parabolic orbits over the past nearly 50~years.}  This result should
have raised a {\small \bf red flag}, as it amounts to a powerful {\small \bf
argument against the nongravitational acceleration in the orbital motion of
`Oumuamua having been induced by the sublimation of water ice}.  It is unfortunate
that this major finding is noted by Micheli et al.\ only peripherally in the text
and Table~1, but not explicitly in the executive summary.\footnote{Micheli et al.'s
brief note on an improved model that is hugely asymmetric relative to perihelion
is, as far as I noted, accompanied in their paper by no comment on the physical
significance of $\Delta T$, which I~believe is in this case very probably none
at all.}

Micheli et al.'s nominal outgassing model requires --- precisely for the
reason just elaborated upon --- a major contribution from carbon monoxide, whose
sublimation rate varies, unlike that of water ice, as $r^{-2}$ in~the~range of
relevant heliocentric distances, thus complying with the observed law of variation. 
Since Trilling et al.'s (2018) results, based on the nondetection of `Oumuamua
with the Spitzer Space Telescope, demonstrate conclusively that the nongravitational
acceleration could under no circumstances be a corollary of the sublimation
of carbon dioxide or carbon monoxide, I focus in this paper on providing
a wealth of arguments to support the notion that the acceleration cannot in
fact be a product of the sublimation of water ice either.

\begin{table*}[t]
\vspace{-4.2cm}
\hspace{-0.52cm}
\centerline{
\scalebox{1}{
\includegraphics{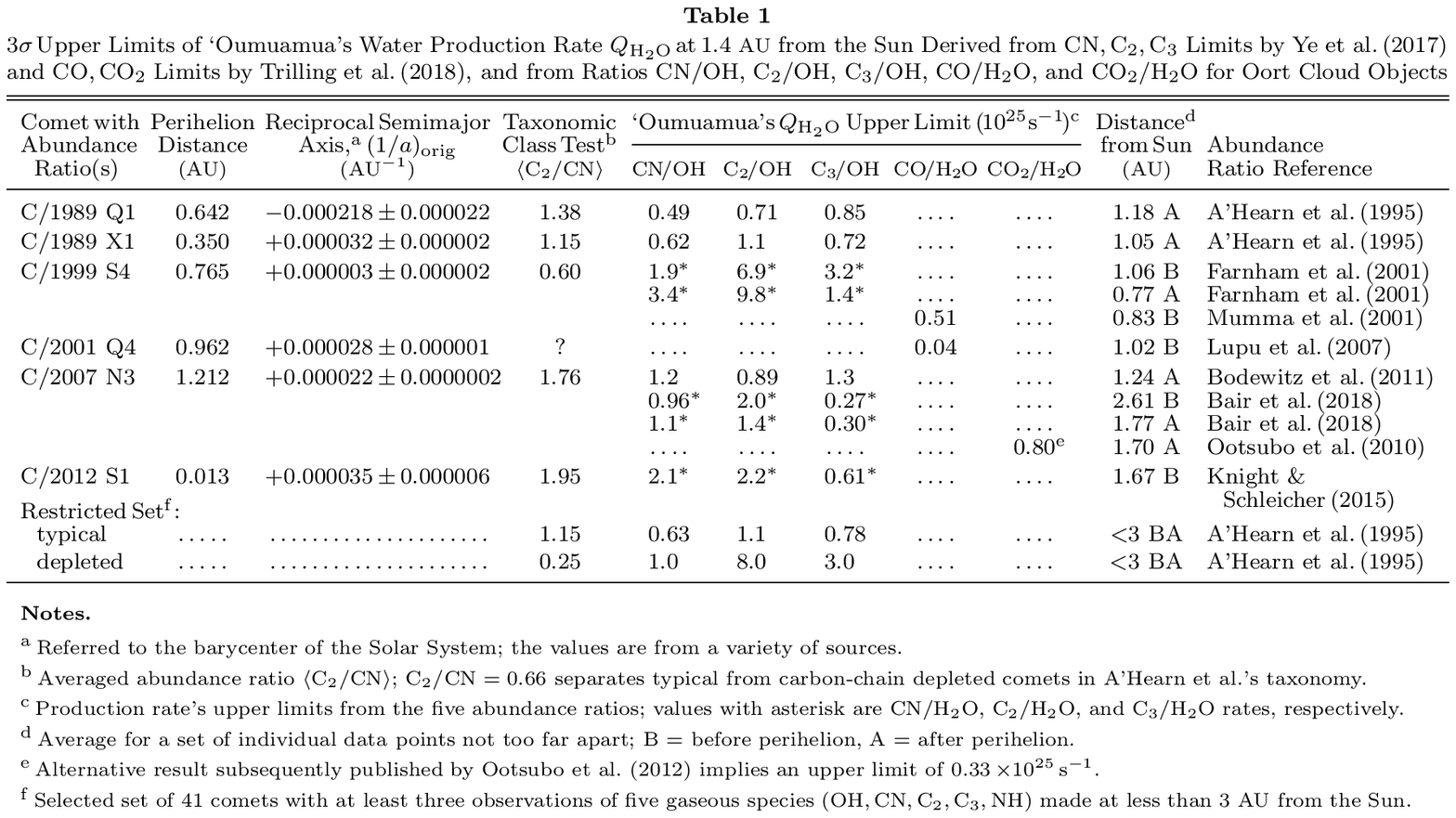}}}
\vspace{-14.7cm}
\end{table*}

\section{Constraints from Gas Abundance Ratios}

A meaningful upper limit of the water production~rate for `Oumuamua can be estimated
from the upper limits of the production rates of CN, C$_2$, C$_3$, CO, and CO$_2$,
summarized in Section~1, combined with the known production rate ratios CN/OH,
C$_2$/OH, C$_3$/OH (or directly CN/H$_2$O, C$_2$/H$_2$O, C$_3$/H$_2$O),\footnote{The
production rates of OH and H$_2$O are assumed equal, the other water dissociation
and ionization products being neglected.} CO/H$_2$O, and CO$_2$/H$_2$O for other
appropriately chosen objects.  Considering that `Oumuamua could have been part of
an Oort Cloud of another stellar system (e.g., Do et al.\ 2018), the members
of the Solar System's Oort Cloud appear to be the closest analogs available for
assessing `Oumuamua's composition properties.

Table 1 lists six Oort Cloud comets used in~this~exercise.~Columns~1--3,~5--9,
and~11~are~\mbox{self-explanatory},~while the fourth shows the test of A'Hearn
et al.'s (1995) taxonomic class, $\langle {\rm C}_2/{\rm CN}\rangle$, an average
of the ratio over the available data, identifying a member of the class~of typical
objects when it is $\geq$0.66, or of the class~of carbon-chain depleted
objects otherwise.  Of the six entries, only C/1999~S4 was carbon-chain depleted,
while there are no data to classify C/2001~Q4.  For comparison, I also include
`Oumuamua's water production rates derived from A'Hearn et al.'s average
values for the groups of typical (70~percent of the total) and depleted objects
in the restricted set, which contains the data obtained for the 41~comets that
were well observed at heliocentric distances of less than 3~AU.  The averages
of the restricted set's~typical ratios amount to:\
\mbox{CN/OH\,=\,0.0032\,$\pm$\,0.0013}, \mbox{C$_2$/OH = 0.0036\,$\pm$\,0.0017},
\mbox{C$_3$/OH = 0.00026\,$\pm$\,0.00017}; for the depleted ones they
are:\,\mbox{CN/OH\,=\,0.0020\,$\pm$\,0.0007}, \mbox{C$_2$/OH = 0.0005\,$\pm$\,0.0004},
\mbox{C$_3$/OH = 0.00007\,$\pm$\,0.00004}.  Column 10 specifies which part of the
orbit do the data come from, listing an average heliocentric distance~and the
orbital branch, B if before or A if after perihelion.

The 3$\sigma$ upper limits of `Oumuamua's production rate of water in Table~1 apply
to a heliocentric distance of about 1.4~AU, at which Ye et al.'s (2017) observations
were made and to which Micheli et al.\ (2018)~\mbox{refer}~their values as well.
The production rates derived~from the Spitzer Telescope observations (Trilling et
al.\ 2018), were multiplied by a factor of four, implied~by~\mbox{Marsden} et
al.'s (1973) Style II nongravitational law, to get converted to 1.4~AU.  The
results indicate that the H$_2$O upper limit is much tighter than
$\sim$10$^{27}$\,molecules~s$^{-1}$;~five comets~consistently show it to be
more than two orders of~\mbox{magnitude} lower, {\small \bf 3~to~\mbox{6{\boldmath
$\times\:\!\!$10$^{24}$}molecules~s{\boldmath $^{-1}\!$}}},~while~the~sixth~comet,
C/2001~Q4, offers a limit by yet another order of magnitude lower, {\small \bf
\mbox{4{\boldmath $\times\:\!\!$}10{\boldmath $^{23}$}\,molecules s{\boldmath
$^{-1}$}}}.  The data from the restricted set (A'Hearn et al.\ 1995), although of
lesser relevance, suggest nearly equally low limits, including the class of
carbon-chain depleted objects that generally provide somewhat higher, and less
robust, limits.

In addition to the observationally established deficit of at least 3--4 orders of
magnitude in the production rate of carbon monoxide (Trilling et al.\ 2018) and
next~to the contradiction in the nongravitational law (Section~2), {\small \bf
`Oumuamua's sublimation rate of water ice derived from five gas-abundance ratios
is at least 1--2 orders of magnitude short of the production level required
by the model interpreting the observed  nongravitational acceleration as an
outgassing-driven effect}.~This~major {\small \bf disparity} increases to {\small \bf
at least 2--3 orders of magnitude for the pendulum-rotation model}~proposed 
by Seligman et al.\ (2019).

\begin{table*}
\vspace{-4.22cm}
\hspace{-0.51cm}
\centerline{
\scalebox{1}{
\includegraphics{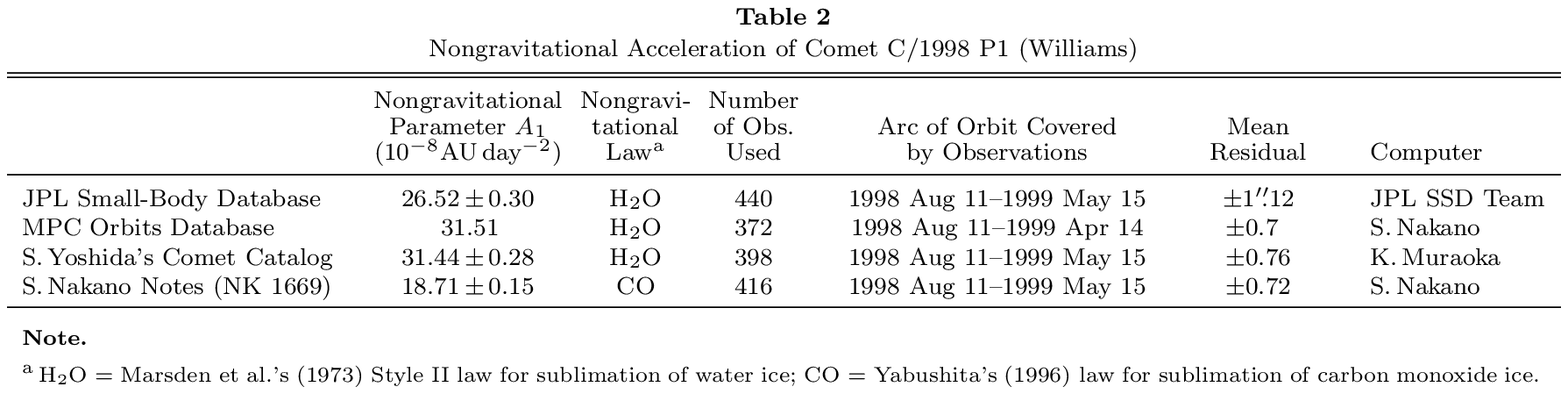}}}
\vspace{-20.2cm}
\end{table*}

\section{Comparison with Nongravitational Effects\\in the Motions of Other Comets}
Normalized to 1 AU from the Sun, the nongravitational acceleration detected
{\vspace{-0.04cm}}by Micheli et al.\ (2018)~in~the orbital motion of `Oumuamua is
\mbox{24.55\,$\times\:\!  $10$^{-8}$\,AU day$^{-2}$}; it is directed radially
away from the Sun and varies as an inverse second or first power of heliocentric
distance.

To determine whether the magnitude of `Oumuamua's nongravitational acceleration
fares well in comparison with the outgassing-driven effects in the motions
of long-period comets,\footnote{The radial component of the nongravitational
acceleration has a different meaning for linked apparitions of the short-period
comets, given the symmetric law employed (Sekanina 1993).} I researched four online
comet orbit catalogs:\ {\vspace{-0.03cm}}the {\it Jet Propulsion Laboratory\/}'s
(JPL) Small-Body Database;\footnote{See {\tt
https://ssd.jpl.nasa.gov/sbdb\_query.cgi}.} {\vspace{-0.03cm}}the {\it
Minor Planet Center\/}'s (MPC) Orbits/Observations{\vspace{-0.03cm}}
Database;\footnote{See {\tt https://minorplanetcenter.net/db\_search};\,also\,the\,now-outdated printed catalogue by Marsden \& Williams (2008).}
S.\ Yoshida's cometary catalogue;\footnote{See
{\tt http://www.aerith.net/comet/catalog/index-code.html}.} and S.~Nakano's
series of circulars.\footnote{See {\tt http://www.oaa.gr.jp/$\sim$oaacs/nk.htm}.}
The net result of this extensive search was that the {\small \bf number of objects}
with a radial nongravitational acceleration comparable to, or greater than,
`Oumuamua's acceleration {\small \bf was exactly one --- C/1998~P1
(Williams)}, whose perihelion distance was 1.146~AU.  However,
as illustrated in Figure~1, the findings are complicated by the different
nongravitational laws employed.  From the orbit-determination details, listed in
Table~2, it follows that the {\it integrated effect\/} of the nongravitational
acceleration for C/1998~P1 was in fact smaller than for `Oumuamua.  The parameter
$A_1$ for C/1998~P1 slightly exceeded that for `Oumuamua only when the water ice
sublimation law was forced through the observations (the JPL, MPC, and Yoshida
entries in Table~2).  When this law was replaced with Yabushita's (1996)
carbon-monoxide sublimation law (the Nakano Notes entry), which better fits
the astrometric observations and closely matches the inverse square power law of
heliocentric distance used by Micheli et al.\ (2018) for `Oumuamua,{\vspace{-0.04cm}}
the parameter $A_1$ for C/1998~P1 dropped to merely $\frac{3}{4}$ `Oumuamua's
acceleration.  The two objects also differed dramatically in their appearance,
`Oumuamua looking stellar, the comet progressively more diffuse,
displaying no nuclear condensation at all (\mbox{DC = 0}) when observed visually
by Hale (2000) in March 1999, before disappearing on its way out.  In any case,
{\small \bf `Oumuamua's orbital anomaly does not compare
well with the outgassing-driven effects in the motions of long-period comets,
exceeding even the largest among them by a sizable margin}.

At the other extreme, the dwarf members of the Kreutz sungrazing system are
subjected to much higher nongravitational accelerations than is `Oumuamua's shortly
before they sublimate away in the Sun's corona (Seka\-nina \& Kracht 2015), a
scenario that is here irrelevant.  Accelerations comparable to that of
`Oumuamua are common among short-lived companions of the split comets, which
are also of diffuse appearance and disintegrate on a time scale of weeks near
1~AU from the Sun (Sekanina 1982), also unlike `Oumuamua.~In summary, there appears
to be {\small \bf no object
whose outgassing-driven nongravitational effects and appearance both resemble
 `Oumuamua's}.

\begin{figure}[b]
\vspace{-1.68cm}
\hspace{0.64cm}
\centerline{
\scalebox{0.645}{
\includegraphics{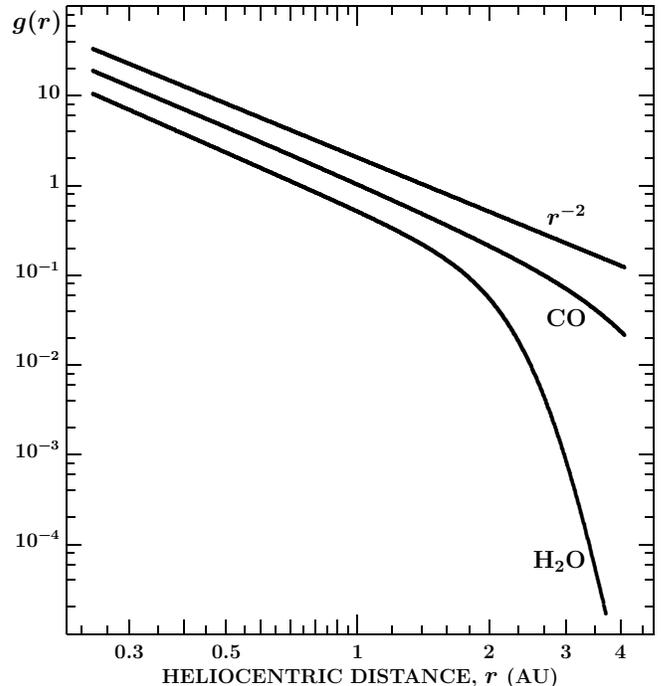}}}
\vspace{-8.19cm}
\caption{Comparison of three nongravitational laws $g(r)$.  The H$_2$O curve is
the standard Style II law by Marsden et al.\ (1973), describing the sublimation
of water ice.  The CO curve is the~carbon monoxide law by{\vspace{-0.05cm}}
Yabushita (1996), applied by Nakano to comet C/1998~P1.  The $r^{-2}$ curve is the
inverse square power law~of~heliocentric distance employed by Micheli et al.\
(2018) for the nongravitational acceleration of `Oumuamua.  The three curves are
essentially equivalent up to about 1~AU from the Sun;~the~H$_2$O~law~then
diverges rapidly. The curves are vertically shifted for~clarity.{\vspace{0.1cm}}}
\end{figure}

\section{Water Sublimation Modeling:\\The Inconsistencies}
The next step in my pointing out the weaknesses of the interpretation of `Oumuamua's
nongravitational acceleration as a phenomenon driven by the sublimation of water
ice is to demonstrate the implications, which are extreme.  To start with, I
consider a piece of matter of unit cross-sectional area, length $\ell$, and
uniform bulk density $\rho$, which includes at one end a layer of water ice
sublimating at a constant rate of $Z$ molecules per unit area per second with a
speed $v$ isotropically into a hemisphere whose boundary plane is perpendicular
to the length $\ell$, thus imparting to the rest of the body an acceleration $\Psi$.
The conservation of momentum law dictates an equality
\begin{equation}
\rho \ell \Psi = \langle Z m v \rangle ,
\end{equation}
where $m$ is the mass of a water molecule and $\langle Zmv \rangle$~is the component
of the momentum change $Zmv$ in the~direction parallel to the body's length,
\mbox{$\langle Zmv \rangle = \frac{2}{3} Zmv$};~the components in all other
directions cancel out.

Equation (1) represents the condition that the length $\ell$ of the body should,
as a function of the bulk density, satisfy in order to fit the acceleration under
the constraints.  Given that `Oumuamua's nongravitational acceleration varies as
an inverse square of heliocentric distance,~it equals~\mbox{$\Psi = \:\!\!12.5
\:\!\!\times \! 10^{-8}\:\!$AU day$^{-2} \:\!\!= 2.51 \:\!\! \times \! 10^{-4}\:\!
$cm s$^{-2}$}~at 1.4~AU from the Sun.  The standard equation of energy balance,
applied to the solar flux incident on a Sun-facing planar surface of water~ice,
of an albedo of 0.1 and emissivity of 0.9, which is spent on the sublimation of
the ice and on the black-body energy reradiation, provides at a heliocentric
distance of 1.4~AU the sublimation~rate~of \mbox{$Z = 6.9 \times \:\!\! 10^{17}
$\,molecules cm$^{-2}$ s$^{-1}$ and the temperature} of \mbox{$T = 199.8$\,K}.
In the absence of dust emission, the sublimation {\vspace{-0.07cm}}velocity equals,
according to Probstein~(1969), the speed of sound, \mbox{$v_{\rm s} = \sqrt{k_{\rm
B} \gamma_0 T/m}$}, where $k_{\rm B}$ is the Boltzmann constant and $\gamma_0$ is the
adiabatic index for water, amounting to 1.33.  With the numerical values fitted in,
\mbox{$v_{\rm s} = 3.50 \times \:\!\! 10^4$\,cm s$^{-1}$}, yielding for the product
of $\rho \ell$:
\begin{equation}
\rho \ell = 1920 \; {\rm g} \; {\rm cm}^{-2}.
\end{equation}
For \mbox{$\rho \simeq 1$ g cm$^{-3}$}, for example, the length \mbox{$\ell \simeq
20$ m},~a~dimension significantly shorter than the expected characteristic size of
`Oumuamua, in spite of the sublimation rate being maximized by assuming the normal
incidence of the solar flux and neglecting the fraction of the solar energy that
penetrates into the interior of the body.  Equation~(2) indicates that the length
could be increased only by {\small \bf lowering the bulk density}.

This elementary scenario is now expanded in two steps.  First, in order to gain
some insight, I assume a crude, spherical approximation, for which the
nondetection of `Oumuamua by the Spitzer Space Telescope implies a
diameter of less than 140~m at a visual geometric albedo of \mbox{$p_V = 0.1$}
or corresponding diameters at other albedo values according to Trilling et al.\
(2018).  It should be remembered, however, that there are three conditions that
need to be satisfied by a self-consistent solution simultaneously and that {\small
\bf the compliance} with all of them still {\small \bf fails to remove the
red-flag problem of the water sublimation law}, brought up to the reader's
attention in Section~2.

The first is the condition dictated by the conservation of momentum law, which
now requires the determination of the contributions from all sublimating surface
elements on the sunlit hemisphere to the component~in the radial direction, that is,
along the Sun-nucleus line.  The sublimation rate at the subsolar point, $Z_0$, is
equal to the sublimation rate $Z$ used above in the one-dimensional scenario.  The
sublimation rate $Z(\theta)$ at any other point on the sunlit hemisphere, at which
the normal to the spherical surface makes an angle $\theta$ with the direction to
the Sun, is lower because the incident solar flux amounts to only a $\cos \theta$
fraction of the solar flux at the subsolar point.  I address this issue fully
below~in~Section~6.~For the purpose of the spherical approximation~I~assume~that
\mbox{$Z(\theta) = Z_0 \cos \theta$} for any $\theta$, which, as will be seen
later, moderately exaggerates the actual sublimation rate.  The sublimation
velocity $v_{\rm s}$ decreases with increasing angle $\theta$ very slowly because
of the nearly constant temperature, dropping at \mbox{$\theta = 85^\circ$} to 0.94
the velocity at \mbox{$\theta = 0^\circ$},~an~effect that is neglected in this
first step.~The~\mbox{sublimation} rate from an annulus located between $\theta$ and
\mbox{$\theta \!+\!d\theta$}~on~the sunlit half of the spherical nucleus of
diameter~$D$~equals{\vspace{-0.01cm}}

{\noindent}\mbox{$\frac{1}{2} \pi D^2 \:\!\!Z_0 \sin \theta \cos \theta
\,d\theta$.$\;$A momentum per unit time trans-} ferred to the nucleus in the antisolar
direction amounts to \mbox{$\frac{1}{2} \pi D^2 (\frac{2}{3} Z_0 m v_{\rm s})
\sin \theta \cos^2 \!\theta\,d\theta$}.  {\vspace{-0.09cm}}Summing up this expression
over the sunlit hemisphere yields $\frac{1}{9} \pi D^2 Z_0 m v_{\rm s}$ and the
conservation of momentum law requires that
\begin{equation}
{\textstyle \frac{1}{6}} \pi \rho D^3 \Psi = {\textstyle \frac{1}{9}} \pi D^2
 Z_0 m v_{\rm s},
\end{equation}
where on the left is the product of the nucleus' mass and acceleration.  This
expression simplifies to
\begin{equation}
\rho D \Psi = {\textstyle \frac{2}{3}} Z_0  m v_{\rm s},
\end{equation}
which is Equation (1) when \mbox{$\ell = D$}, \mbox{$\langle Z \rangle \!=\!
{\textstyle \frac{2}{3}} Z_0$}, and \mbox{$v \!=\! v_{\rm s}$}.

The second condition is the requirement that the sublimation rate integrated
over all surface elements be equal to the production rate $Q_{{\rm H}_2{\rm O}}$.
Integrating, under the employed approximation, the sublimation rate from the
annulus between $\theta$ and \mbox{$\theta \!+\! d\theta$} over the entire
sunlit hemisphere gives
\begin{equation}
Q_{{\rm H}_2{\rm O}} = {\textstyle \frac{1}{4}} \pi D^2 Z_0 .
\end{equation}

The third condition equates the product of the projected area
of the nucleus and the geometric albedo $p_V$ with the
intrinsic brightness:
\begin{equation}
{\textstyle \frac{1}{4}} \pi D^2 p_V = X_0.
\end{equation}

The quantities $\Psi$, $Z_0$, $v_{\rm s}$, and $Q_{{\rm H}_2{\rm O}}$ in the
conditions (4) and (5) are heliocentric-distance dependent, while the condition
(6) contains constants.  In a previous paper (Sekanina 2019b) I remarked that,
based on Drahus et al.'s (2018) data, the projected area of{\vspace{-0.04cm}}
`Oumuamua at peak brightness equaled \mbox{0.002/$p_V$ km$^2$}.  With the
light-curve amplitude between 6:1 and 11:1, the mean projected area over a
rotation cycle implies for the constant on the right-hand side of Equation~(6)
\mbox{$X_0 = 0.0011$ km$^2$}.  At \mbox{$p_V = 0.1$}, Equation~(6) implies
\mbox{$D = 118$ m}, consistent with Trilling et al.'s (2018) upper limit of
140~m.

Continuing to refer the variable quantities to a heliocentric
distance of 1.4~AU, I find from Equation~(4)
\begin{equation}
\rho D = 1920 \; {\rm g \; cm}^{-2} \! ,
\end{equation}
which copies Equation (2), replacing only $\ell$ with $D$.

Adopting \mbox{$\rho = 0.5$ g cm$^{-3}$} as a typical bulk density~of~a cometary
nucleus, this relation implies \mbox{$D \simeq 38$ m}, a value about a factor
of three smaller than derived above for \mbox{$p_V = 0.1$} from Equation~(6).
Keeping the sublimation quantities in Equation~(4) fixed at this point and avoiding
unrealistically high albedos, this disparity can only be removed, as in Equation~(2),
{\vspace{-0.04cm}}by lowering the bulk~density, in this case down to \mbox{0.16
g cm$^{-3}$}.  Next, by combining the conditions (4) through (6), I examine the
constraints imposed by the water production rates established in Section~3 ---
the softer limit of \mbox{6\,$\times\:\!$10$^{24}$\,molecules s$^{-1}$} and the
tighter limit of \mbox{4\,$\times\:\!$10$^{23}$\,molecules~s$^{-1}$}:\ (i)~on $Z_0$
from an expression that is independent of $D$ and $\rho$; (ii)~on $D$ from an
expression that is independent of $\rho$; and (iii)~on $\rho$ from an expression
that is independent of $D$.

The first constraint is obtained by eliminating $\pi D^2$ from Equations~(5)
and (6):
\begin{equation}
Z_0 = \frac{p_V Q_{{\rm H}_2{\rm O}}}{X_0} .
\end{equation}
The second constraint is acquired directly from Equation~(5):
\begin{equation}
D = \frac{2}{\sqrt{\pi}} \sqrt{\frac{Q_{{\rm H}_2{\rm O}}}{Z_0}}.
\end{equation}
One can formulate two independent constraints on the bulk density.~By
eliminating $D$ from (4) and (6) and~then inserting for $Z_0$ from (8)
one obtains
\begin{equation}
\rho = \frac{\sqrt{\pi} m v_{\rm s} Q_{{\rm H}_2{\rm O}}}{3 \Psi} \! \left(
 \! \frac{p_V}{X_0} \! \right)^{\!\!\frac{3}{2}} \!\! ,
\end{equation}
while by eliminating $D$ from (4) and (5),
%
\begin{equation}
\rho = \frac{\sqrt{\pi} m v_{\rm s}}{3 \Psi}
 \sqrt{\frac{Z_0^3}{Q_{{\rm H}_2{\rm O}}}} \, .
\end{equation}
The upper limit of $Q_{{\rm H}_2{\rm O}}$ offers, in Equation~(8), an upper
limit of $Z_0$ (to be compared with $Z_0$ from the equation of energy balance);
in Equation~(9), an upper limit of the nucleus' diameter (to be compared with the
photometric value of 118~m); and in Equation~(10), an upper limit of the bulk
density; besides, it provides, in Equation~(11), a {\it lower\/} limit of the
bulk density.

Inserting the numerical values, the problems emerge immediately; inserted
into Equation~(8), the two upper limits of the water production rate imply the
sublimation rates of, respectively,{\vspace{-0.05cm}}
\begin{eqnarray}
Z_0 & < & 5.5 \!\times \!\!10^{17} \!p_V \!\!=\! 0.55 \!\times \!\! 10^{17} 
{\rm molecules\:cm}^{-2}{\rm s}^{-1}{\hspace{-0.03cm}}{\rm ;\;or}\nonumber\\[0.04cm]
Z_0 & < & 0.36 \!\times \!\!10^{17} \! p_V \!\!=\! 0.036 \!\times \!\!10^{17} 
 {\rm molecules\:cm}^{-2}{\rm s}^{-1}
\end{eqnarray}
with the geometric albedo of \mbox{$p_V = 0.1$}; they are nowhere near the expected
\mbox{$Z_0 = 6.9 \times\:\!\!  10^{17}$\,molecules cm$^{-2}$ s$^{-1}$}.  The
constraint on the diameter of the object from (9)~is likewise
unacceptable{\vspace{-0.05cm}}
\begin{eqnarray}
D & < & 33 \;{\rm meters; \; or} \nonumber \\[0.01cm]
D & < & 8.6 \;{\rm meters} .
\end{eqnarray}
The constraints on the bulk density are even more extreme; from
Equation~(10){\vspace{-0.05cm}}
\begin{eqnarray}
\rho & < & 0.41 \,p_V^{\frac{3}{2}} = 0.013\;{\rm g\;cm}^{-3};\;{\rm or}
 \nonumber\\[-0.05cm]
\rho & < & 0.027 \,p_V^{\frac{3}{2}} = 0.00085 \;{\rm g \; cm}^{-3} \!,
\end{eqnarray}
and from Equation~(11)
\begin{eqnarray}
\rho & > & 0.58 \;{\rm g \; cm}^{-3}; \, {\rm or} \nonumber \\[-0.05cm]
\rho & > & 2.23 \;{\rm g \; cm}^{-3} \! .
\end{eqnarray}
The upper limit of the bulk density is orders of magnitude below
its lower limit, confirming that there is no way to make
the nongravitational acceleration consistent with the upper limit of
the water production rate under the given conditions.

The clue to removing the inconsistencies in modeling the production of
water is by eliminating the discrepancy between the sublimation rates
per unit area, $Z_0$, determined from the energy balance equation on
the one hand and by the inequalities (12) on the other hand.
This effort is equivalent to making sure that the
upper and lower limits of the bulk density equal each other, as
both lines of attack lead to the crucial equation (8).  To satisfy the
tighter of the inequalities (12), the theoretical water sublimation rate
per unit surface area of \mbox{$Z_0 = 6.9 \:\!\! \times \! 10^{17}$
molecules cm$^{-2}$\,s$^{-1}$} ought to be reduced by a factor of more
than $\sim$200 for \mbox{$p_V \simeq 0.1$}.  This problem is solved by
admitting that the {\small \bf sublimation of water ice proceeds from
less than 0.5~percent of the nucleus' surface}.  However, reducing $Z_0$
so dramatically has an effect on the condition in Equation~(7), which now
reads
\begin{equation}
\rho D < 9.6 \; {\rm g \;cm}^{-2}.
\end{equation}
A diameter of 118 m then implies an {\small \bf upper limit~of~only {\boldmath
$\sim$}0.0008~g~cm{\boldmath $^{-3}$} for the bulk density of `Oumuamua}, thus
making it an extremely porous object of exceptional, fluffy-like morphology,
{\small \bf a lookalike of the piece of debris of a dwarf interstellar comet,
whose nongravitational acceleration was} proposed to have been {\small \bf
driven by solar radiation pressure} (Sekanina 2019a).

\section{Water Sublimation Modeling Refined}
In step two of this exercise I remove the assumption of the spherical shape
of `Oumuamua, which is contrary to observational evidence, and replace the
approximate expressions for the water sublimation rate and temperature
variations with the zenith distance of the Sun with more rigorous expressions,
before addressing the three conditions represented in step one by
Equations~(4)~to~(6).

The shape of `Oumuamua is now approximated by a right circular cylinder,
whose bases have a diameter of $D$ and the side a length of $\ell$.
The object is either~a~disk (\mbox{$D \gg \ell$}), equivalent to an
oblate-spheroid model;\,or a rod (\mbox{$D \ll \ell$}), equivalent to
a prolate-spheroid model.  In the existing terminology, `Oumuamua is
being referred to as a pancake-shaped or cigar-shaped object, respectively.
In the previous paper (Sekanina 2019b) I noted that only the pancake-like
version was consistent with the nondetection by the Spitzer Telescope
(Trilling et al.\ 2018), the cigar version's dimensions exceeding the
3$\sigma$ limits.

\begin{table}[t]
\vspace{-3.93cm}
\hspace{4.23cm}
\centerline{
\scalebox{1}{
\includegraphics{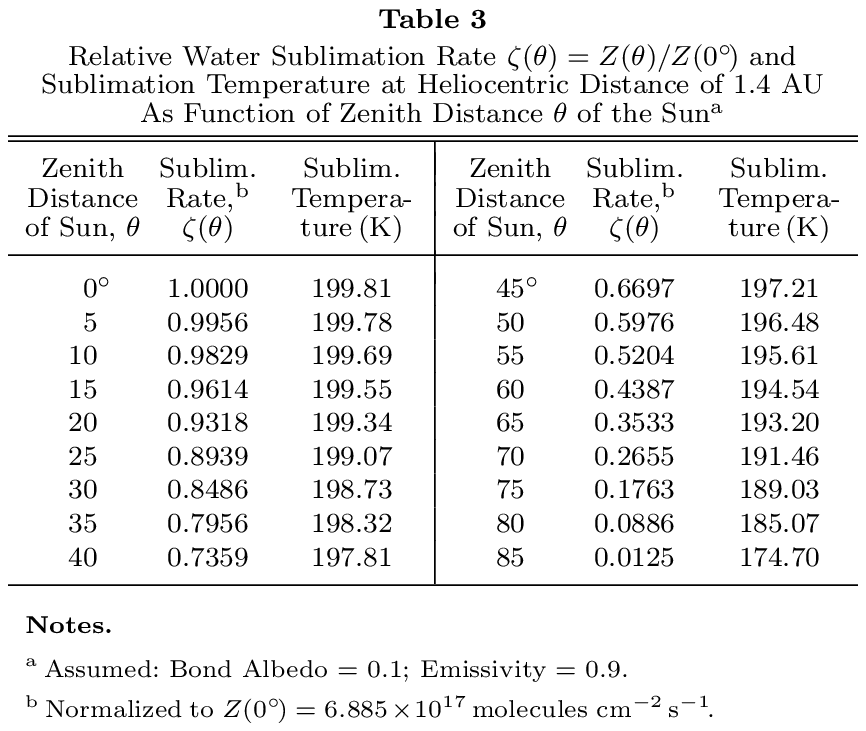}}}
\vspace{-18.2cm}
\end{table}

In order to verify or disprove the conclusions of the previous section,
I deduce the expressions for the momentum change exerted on the object
in the antisolar direction to replace Equation~(4); for the overall
water production from the object to replace Equation~(5); and for its
projected area to replace Equation~(6); by averaging over all orientations
of `Oumuamua relative~to~the{\nopagebreak} Sun.  For this purpose, the dependence
of the water~\mbox{sublimation}~rate per unit surface area, $Z(\theta)$, and the
sublimation temperature, $v_{\rm s}(\theta)$, on the Sun's zenith distance
$\theta$ have been computed by incorporating $\theta$ into the energy
balance equation; for a heliocentric distance of 1.4~AU the results are
presented in Table~3, in which $\zeta(\theta)$ is the relative sublimation
rate per unit surface area at the Sun's zenith distance $\theta$, expressed
in units of the subsolar sublimation rate \mbox{$Z(0^\circ\!\!\:)$}, so that
\mbox{$Z(\theta) = Z(0^\circ\:\!\!) \!\cdot\! \zeta(\theta)$}.

Because of the symmetry of the cylindrical model, only one of the two bases needs
to be considered~and~only~one half of the side.  The position of the Sun is
described by the angle $\theta$ its direction subtends with the axis of the
cylinder, which is the Sun's zenith distance viewed from the base.  The angle
along the circumference~of~the~base, $\phi$, enters only a relation for the
cylinder's side, measuring the angular distance from the plane passing
through the Sun and the cylinder's axis.  The sublimation rate at the subsolar
point, $Z(0^\circ\:\!\!)$ in Table~3, is in the following abbreviated as $Z_0$.

I now deal separately with either~of~the~two~parts~of~the momentum effect
exerted by the water~\mbox{sublimation}~from the base and side of the
cylindrical~body.~At~an~\mbox{angle $\theta$}, the
sublimation~rate~from~a~unit~surface~area~of~the~base
is $Z_0 \zeta(\theta)$, so that the contribution from~this~area to~the
momentum change in a direction normal~to~the surface is
\mbox{$\frac{2}{3} Z_0 \zeta(\theta) m v_{\rm s}(\theta)$}, where $m$ is again
the~mass~of~a water
molecule.  This area's contribution to the momentum change
in the antisolar direction is
\mbox{$\frac{2}{3}Z_0 \zeta(\theta) m v_{\rm s}(\theta)\cos\theta$} and the
overall contribution from the base at the Sun's zenith distance $\theta$ is 
\begin{equation}
{\textstyle \frac{1}{4}} \pi D^2 \!\cdot\! {\textstyle \frac{2}{3}} Z_0 \,
 \zeta(\theta) \, m \, v_{\rm s}(\theta) \cos \theta .
\end{equation}

The Sun's position, described by a zenith distance $\theta$ as viewed from the
base of the cylinder, makes a minimum angle of \mbox{$90^\circ \!\!-\! \theta$}
with the surface of its side along the stripe extending in the plane that contains
the Sun and the cylinder's axis.  At all other points of the cylinder's
sunlit side the Sun's zenith distance is in the range between
\mbox{$90^\circ\!\!-\!\theta$} and 90$^\circ$.  The side's peak sublimation
rate from a unit surface area is therefore $Z_0 \zeta(90^\circ\!\!-\!\theta)$
and its contribution to the momentum change in the antisolar direction equals
\mbox{$\frac{2}{3} Z_0 \zeta(90^\circ \!-\! \theta) m v_{\rm s}(90^\circ\!\!
-\!\theta) \sin \theta$}.  A stripe{\vspace{-0.09cm}} of an infinitesimal width
of $\frac{1}{2}D\,d\phi$ and the length $\ell$ between both bases contributes
\begin{equation}
{\textstyle \frac{1}{2}} D\ell\,d\phi \!\cdot\! {\textstyle \frac{2}{3}} Z_0 \,
 \zeta(90^\circ \!\!-\! \theta) \, m \, v_{\rm s}(90^\circ \!\!-\! \theta)
 \sin \theta .
\end{equation}
A similar stripe of infinitesimal width whose normal to the surface of the side
subtends with the Sun an angle $\psi$ and with the plane passing through the Sun
and the cylindrical axis an angle $\phi$ reckoned in this plane in both directions
along the circumference of the base, sublimates at the Sun's zenith angle of
\mbox{$90^\circ \!\!-\! \theta$} at a rate of \mbox{$Z_0 \zeta(\psi)$}, where
the angle $\psi$ depends on $\theta$ and $\phi$ through a relation
\begin{equation}
\cos \psi = \sin \theta \cos \phi.
\end{equation}
This stripe's contribution to the momentum change in the antisolar direction is
\begin{equation}
{\textstyle \frac{1}{2}} D\ell\,d\phi \!\cdot\! {\textstyle \frac{2}{3}} Z_0 \,
 \zeta(\psi) \, m \, v_{\rm s}(\psi) \cos \psi.
\end{equation}
It is noted that \mbox{$\psi = 90^\circ\!\!-\!\theta$} and the expression (20)
becomes identical with (18) when \mbox{$\phi = 0^\circ$}.  Integrating the
contributions to the momentum change in the antisolar direction over the surface
of the sunlit half of the cylindrical side, I find for the Sun's angle $\theta$,
\begin{eqnarray}
 & {\textstyle \frac{1}{3}} D \ell Z_0 m \!
 {\displaystyle \int_{-\frac{\pi}{2}}^{\frac{\pi}{2}}} \!\zeta(\psi) \,
 v_{\rm s}(\psi) \, \cos \psi \,d\phi & \nonumber \\[-0.05cm]
{\rm or} & & \nonumber \\[-0.05cm]
 & {\textstyle \frac{2}{3}} D \ell Z_0 m \! {\displaystyle \int_ {\frac{\pi}{2} -
 \theta}^{\frac{\pi}{2}} \!\! \frac{\zeta(\psi) \, v_{\rm s}(\psi)
 \sin \psi \cos \psi}{\sqrt{ \sin {\mbox{\raisebox{-0.15ex}{$\!^2$}}} \theta \!-\!
 \cos^2 \! \psi }}} \, d\psi & \nonumber \\[-0.05cm]
{\rm or} & & \nonumber \\[-0.05cm]
 & {\textstyle \frac{2}{3}} D \ell Z_0 m \sin \theta \!
 {\displaystyle \int_{0}^{\frac{\pi}{2}}} \!\!  \zeta(\psi) \, v_{\rm s}(\psi)
 \cos \phi \, d\phi. &
\end{eqnarray}

In summary, the momentum change $\dot{\mu}(\theta)$, exerted on `Oumuamua in
the antisolar direction at the Sun's angle $\theta$ is given by adding the
expressions (17) and (21),
\begin{eqnarray}
\dot{\mu}(\theta) & = & {\textstyle \frac{1}{6}} D Z_0 m \!\left[
 {\mbox{\raisebox{0ex}[2ex][3ex]{}}} \pi D \, \zeta(\theta) \, v_{\rm s}(\theta)
 \cos \theta \right. \nonumber \\[0cm]
 & & \! \left. + \, 4  \ell \sin \theta \!\!  \int_{0}^{\frac{\pi}{2}} \!\!\!
 \zeta(\psi) \, v_{\rm s}(\psi) \cos \phi \, d\phi \right] \!.
\end{eqnarray}
Averaging over all angles $\theta$, the mean momentum change in the antisolar
direction, $\langle \dot{\mu} \rangle$, becomes
\begin{eqnarray}
\langle \dot{\mu} \rangle & = & \int_{0}^{\frac{\pi}{2}} \!\!\! \mu(\theta)
 \sin \theta \, d\theta {\nonumber} \\
 & = & {\textstyle \frac{1}{6}} D Z_0 m \!\! \int_{0}^{\frac{\pi}{2}} \!\!
 \left[ {\mbox{\raisebox{0ex}[2ex][3ex]{}}} \pi D \, \zeta(\theta) \,
 v_{\rm s}(\theta) \cos \theta \right. \nonumber \\[-0.05cm]
 & & \! \left.+ \, 4 \ell \sin \theta \!\!
 \int_{0}^{\frac{\pi}{2}} \!\!\! \zeta(\psi) \, v_{\rm s}(\psi) \cos \phi \,d\phi
 \right] \! \sin \theta \, d\theta ,
\end{eqnarray}
where the angle $\psi$ is computed from Equation~(19).  The conservation of momentum
condition now requires that
\begin{equation}
\langle \dot{\mu} \rangle = {\textstyle \frac{1}{4}} \pi \rho D^2 \ell \, \Psi,
\end{equation}
where $\Psi$ is, as before, `Oumuamua's nongravitational acceleration.

The expression for `Oumuamua's production rate of water is derived in a manner
similar to the expression for the momentum change.  For any particular
direction~$\theta$ I~find for the production rate summed up from the sunlit
base and side,
\begin{equation}
Q_{{\rm H}_2{\rm O}}(\theta) = {\textstyle \frac{1}{4}} \pi D^2 Z_0 \,
 \zeta(\theta) + 2 D \ell Z_0 \!\!  \int_{0}^{\frac{\pi}{2}} \!\!\! \zeta(\psi)
 \, d\phi,
\end{equation}
where $\psi$ is again given by Equation~(19).  Averaging over all angles
$\theta$ gives for the water production rate
\begin{equation}
\langle Q_{{\rm H}_2{\rm O}} \rangle = {\textstyle \frac{1}{4}} D Z_0 \!\!
 \int_{0}^{\frac{\pi}{2}} \!\! \left[ \pi D \, \zeta(\theta) \!+\! 8 \ell \!\!
 \int_{0}^{\frac{\pi}{2}} \!\!\! \zeta(\psi) \, d\phi \right] \sin \theta
 \, d\theta.
\end{equation}

Viewed from a direction $\theta$, the sunlit projected area~of `Oumuamua equals
in the cylinder approximation
\begin{equation}
X(\theta) = {\textstyle \frac{1}{4}} \pi D^2 \cos \theta + D \ell \sin
 \theta,
\end{equation}
which, when averaged over a hemisphere, becomes
\begin{equation}
\langle X \rangle = {\textstyle \frac{1}{4}} \pi D \, ({\textstyle \frac{1}{2}} D
 \!+\! \ell ).
\end{equation}
Expression (27) implies the presence of two minima and two maxima per rotation.
The minima take place at \mbox{$\theta = 0^\circ$}~and \mbox{$\theta = 90^\circ$},
the shallower one bounded by the maxima satisfying a condition \mbox{$\tan
\theta = 4\ell/\pi D$}, implied by Equation~(27).  When the cylinder is
a disk (\mbox{$D \gg \ell$}), the shallow minimum is at \mbox{$\theta = 0^\circ$}
and the deep at \mbox{$\theta = 90^\circ$}; when a rod (\mbox{$D \ll \ell$}), it
is the other way around.  For a disk, the peak projected area is~given~by
\begin{equation}
X_{\rm max} = {\textstyle \frac{1}{4}} \pi D^2 \sqrt{ 1 \!+\! \left( \!
 \frac{4 \ell}{\pi D} \! \right)^{\!\! 2}}.
\end{equation}
Since the projected area at the deep minimum now equals \mbox{$X_{\rm min}
= D \ell$}, the ratio \mbox{$X_{\rm max}/X_{\rm min} = {\cal A}$},
determining the amplitude of the light curve, is
\begin{equation}
{\cal A} = \sqrt{ 1 \!+\! \chi^2},
\end{equation}
where \mbox{$\chi = \pi D/4 \ell$}.  When the cylinder is a rod, it
is more convenient to write
\begin{equation}
X_{\rm max} = D \ell \, \sqrt{ 1 \!+\! \left( \! \frac{\pi D}{4 \ell} \!
 \right)^{\!\! 2}}.
\end{equation}
As the projected area at the deep minimum in this case equals \mbox{$X_{\rm
min} = {\textstyle \frac{1}{4}} \pi D^2$}, the ratio ${\cal A}$ is again
{\vspace{-0.03cm}}given by Equation~(30), in which though now \mbox{$\chi =
4 \ell/ \pi D$}, a reciprocal of the expression for the disk.

The amplitudes of the shallow minima are also given by Equation~(30),
but with the expressions for $\chi$ interchanged:\ \mbox{$\chi =
4 \ell/ \pi D$} for a disk, but \mbox{$\chi = \pi D / 4 \ell$} for
a rod.

An interesting byproduct of the disk vs rod scenarios is that for the
same peak projected area, $X_{\rm max}$, and the same amplitude, ${\cal A}$ 
--- and therefore also for the same minimum projected area, $X_{\rm min}$ ---
the volume of the disk is always significantly smaller than the volume of
{\vspace{-0.05cm}}the rod, by a factor of $({\cal A}^2 - 1)^{\frac{1}{4}}$,
which amounts, for example, to 2.43 for \mbox{${\cal A} = 6$:1} used below;
the volume equals
\begin{equation}
{\cal V} = \frac{\sqrt{\pi}}{2} \! \left(\! \frac{X_{\rm max}}{\cal A}
 \! \right)^{\!\! \frac{3}{2}} \!\! ({\cal A}^2 \!-\! 1)^\kappa \!,
\end{equation}
where $\kappa$ equals $\frac{1}{4}$ for the disk but $\frac{1}{2}$ for the rod.

Numerically, a light-curve amplitude ${\cal A}$ of 6:1 implies
\mbox{$D/\ell = 7.5$} for a disk and \mbox{$\ell/D = 4.6$} for a rod.
Similarly, an amplitude of 11:1 is equivalent to \mbox{$D/\ell = 13.9$} for
a disk and \mbox{$\ell/D = 8.6$} for a rod.  The reciprocal ratios indicate
extremely small amplitudes for the shallow minima, not exceeding 0.02~mag,
thus effectively retracting the above statement about two minima and two maxima
--- the projected area at, and in between, these maxima is essentially constant.

Recalling once again `Oumuamua's peak projected area of \mbox{$X_{\rm max} = 0.02$
km$^2$} at \mbox{$p_V = 0.1$} (Sekanina 2019b), based on Drahus et al.'s (2018)
photometric observations, I now used Equations~(29) through (31) to determine
the dimensions of the object's cylindrical (both disk and rod) model.  In order
that the results could directly be compared with the 3$\sigma$ upper limit
determined by Trilling et al.\ (2018), for the same assumed value of the albedo
$p_V$, from the nondetection of `Oumuamua by the Spitzer Space Telescope, I adopted
\mbox{${\cal A} = 6$:1}.  For the {\small \bf disk} version I obtained \mbox{$D =
158$ m, $\ell = 21$ m}, both deep {\small \bf within Trilling et al.'s limits} of
341~m by 57~m, but for~the~rod version I got \mbox{$\ell = 303$ m, $D = 65$ m}, the
{\small \bf rod's diameter failing to fit the limit}.  This outcome demonstrates
that the {\small \bf only plausible solution is the disk}, paralleling the result
of a pancake vs cigar comparison examined in Section~9 of Sekanina (2019b).

In order to facilitate the integration of the expressions for the momentum change
and the production rate, I fit $\zeta(\theta)$ and $T(\theta)$ [needed to determine
$v_{\rm s}(\theta)$], the two quantities given in tabular form, by explicit
functions.  For the dependence of the relative sublimation rate of water ice per
unit surface area on the Sun's zenith distance I find a good approximation to be
\begin{equation}
\zeta(\theta) = \cos \theta - 0.068  \, \theta^{2.1} \! \left(
 {\textstyle \frac{1}{2}} \pi \!-\! \theta \right)^{0.8} \!\! ,
\end{equation}
where $\theta$ is in radians; and similarly, for the sublimation
temperature (in K),
\begin{equation}
T(\theta) = 199.81 - 3.55 \, \theta^{1.8} (1.58 \!-\! \theta)^{-0.5}\:\!\! .
\end{equation}
The first terms on the right-hand side represent in both expressions the
crude approximations, with the correction terms following.  It is noted
that the first terms were employed to approximate $\zeta(\theta)$
{\vspace{-0.02cm}}and $T(\theta)$ in Section~5.~The constant 1.58 in
Equation~(34) is used instead of $\frac{1}{2} \pi$ to avoid the negative,
indeterminate value of $T$ at \mbox{$\theta = 90^\circ$}.  The computed
sublimation temperatures near this limit are immaterial, because the relevant
areas of the surface provide only a minute contribution~to the total momentum
change.

This completes the steps necessary to integrate Equations~(23) and (26) and thus
to examine the refined sublimation model.  The results of the computations are
summarized in Table~4 for both the viable, disk-like scenario and the rejected,
rod-like scenario, which contradicts the Spitzer Space Telescope's nondetection
of `Oumuamua.  The results generally confirm the findings based on the approximate
method used in Section~5, even though the bulk density constraint is now slightly
less unfavorable to the hypothesis of outgassing-driven nongravitational
acceleration:\ the tighter constraint on the water production rate in Section~3
implies that a 3$\sigma$ {\small \bf upper limit on `Oumuamua's
bulk density in the disk case amounts to 0.0015~g~cm}{\boldmath
$^{-3}$} and is still lower in the rod case.

\begin{table}[t]
\vspace{-4.25cm}
\hspace{4.25cm}
\centerline{
\scalebox{1}{
\includegraphics{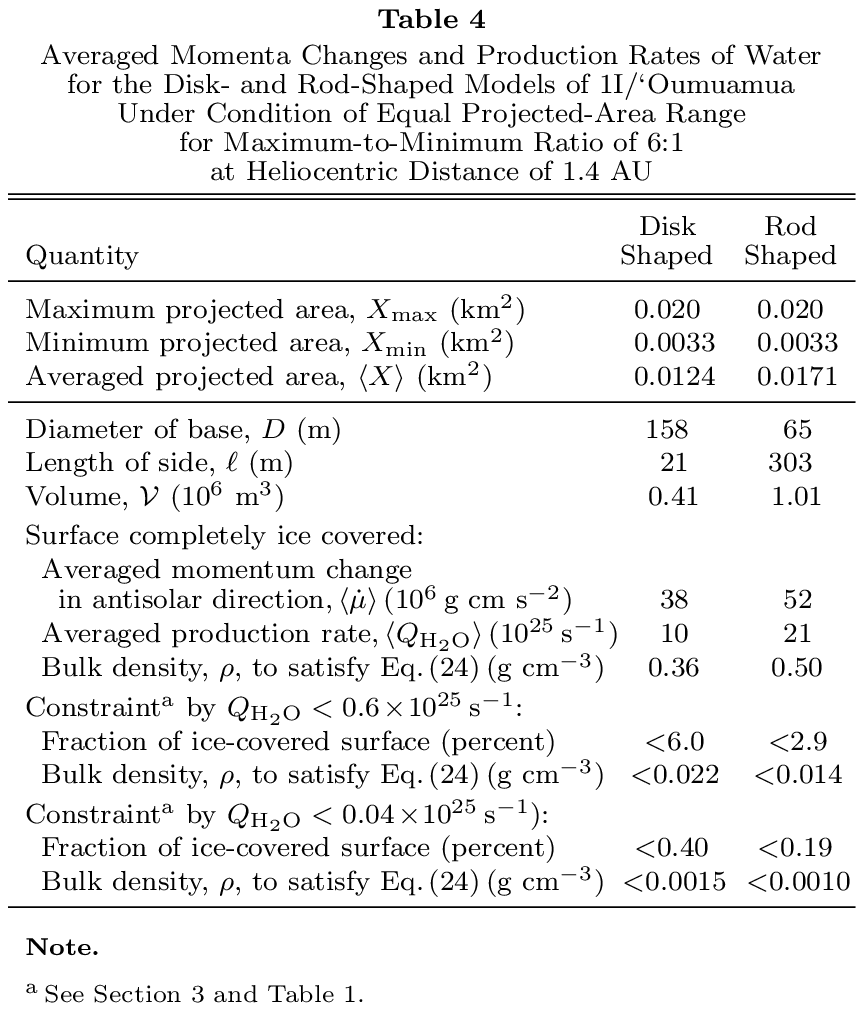}}}
\vspace{-15.2cm}
\end{table}

\section{Discussion}
It has to be remembered that the derived constraints, as 3$\sigma$ upper
limits, exceed in all probability the actual values of the water production
rate and the sublimation-driven momentum change by orders of magnitude, or
that the two are simply zero.  The bulk density should also be substantially lower,
so that --- significantly --- the examination of this hypothesis leads to the
same conclusion as the model in which the nongravitational acceleration is
driven by solar radiation pressure and which requires the object's extremely high
porosity and degree of fluffiness.  Such a model was advocated in my previous
paper based on the argument that `Oumuamua was a piece of inactive debris of
a disintegrated dwarf interstellar comet (Sekanina 2019a).

I already pointed out in Section 5 that the issue of water abundance is separate
from the {\it red-flag\/} problem of the incorrect
law of variation with heliocentric distance that the sublimation of water
ice offers to fit the nongravitational acceleration.  Another problem, a
dramatically different appearance of `Oumuamua and comets with similarly
high nongravitational accelerations, was commented on in Section~4.

\begin{table}[t]
\vspace{-4.25cm}
\hspace{4.21cm}
\centerline{
\scalebox{1}{
\includegraphics{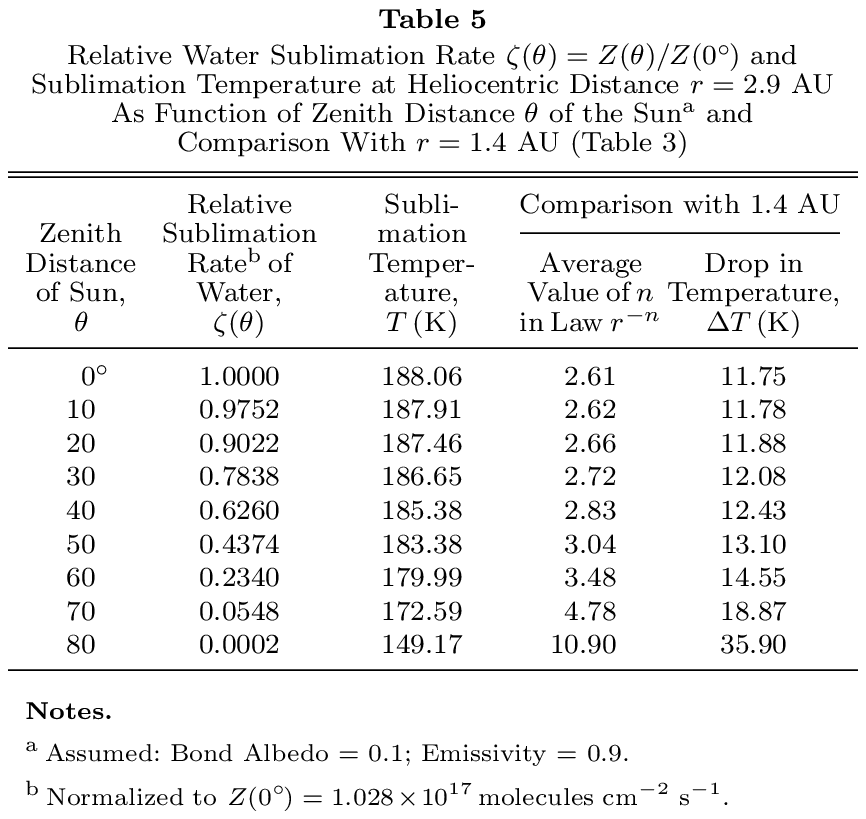}}}
\vspace{-17.2cm}
\end{table}

The red-flag problem of the sublimation law is important enough that it
should be examined in the framework of the cylindrical model for `Oumuamua.
In Section~6 and Tables~3 and 4 I listed the information that this model
provided for the heliocentric distance of 1.4~AU, at which the upper limits
of the production rates of CN, C$_2$, and C$_3$ were obtained by Ye et al.\
(2017) and to which the upper limits of CO$_2$ and CO by Trilling et al.\
(2018) could be converted with sufficient accuracy.  Since `Oumuamua was
discovered only about a week earlier, the heliocentric distance of 1.4~AU
characterizes the period of time when the object was at its brightest.  In order
to appreciate the presumed activity changes over nearly the entire orbital
arc over which `Oumuamua was under observation, I list in Table 5 the
sublimation data, arranged similarly to Table~3, for a heliocentric distance
of 2.9~AU that corresponds to the last observations with the Hubble Space
Telescope in early January 2018.  The last two columns allow comparison of
the expected decrease in both the sublimation rate and temperature over the
period of more than two months that the range of heliocentric distances from
1.4 to 2.9~AU represents.  The issue of the sublimation law is addressed in
column 4, which shows the inverse power $n$ of heliocentric distance $r$, with
which the sublimation rate is expected to drop between 1.4 and 2.9~AU as a
function of the Sun's zenith distance $\theta$.  At the subsolar point the
rate varies as $r^{-2.6}$, but it becomes progressively steeper with increasing
$\theta$, reaching $r^{-3}$ at 50$^\circ$ and nearly $r^{-5}$ at 70$^\circ$, in
contrast with the rate of $r^{-1}$ to $r^{-2}$ established by Micheli et al.\
(2018).  The last column of Table~5 shows that the sublimation temperature
dropped by about 12\,K between 1.4 and 1.9~AU near the subsolar point but
much more at large zenith distances.

In order to examine the integrated effect of the distri\-bution of the sublimation
rate and temperature over the cylindrical model of `Oumuamua, I now applied
the procedure developed in Section~6 to the heliocentric distance of 2.9~AU.
It was necessary to replace the expressions (33) for $\zeta(\theta)$ and
(34) for $T(\theta)$ with new ones; it turned out that their general form
was applicable, with only most of the constants having different values.
The integration showed that for the surface completely covered with water
ice the averaged momentum changes $\langle \dot{\mu} \rangle$ and production
rates $\langle Q_{{\rm H}_2{\rm O}} \rangle$ equaled at 2.9~AU from the Sun:
\begin{eqnarray}
\langle \dot{\mu} \rangle & = & \:\! 4.4 \:\!\! \times \! 10^6 \: {\rm g\;cm}\:{\rm
 s}^{-2} {\hspace{1.27cm}} {\rm for \; disk,} \nonumber \\[-0.05cm]
 & = & \:\!6.1 \:\!\!\times\! 10^6 \: {\rm g\;cm}\:{\rm s}^{-2} {\hspace{1.27cm}}
 {\rm for \; rod,} \nonumber \\[0.05cm]
\langle Q_{{\rm H}_2{\rm O}} \rangle & = & \:\! 0.94 \:\!\! \times \! 10^{25} \:
 {\rm molecules\;s}^{-1} \;\; {\rm for \; disk,} \nonumber \\[-0.05cm]
 & = & \:\!2.03 \:\!\!\times\! 10^{25}\:{\rm molecules\;s}^{-1} \;\; {\rm for \; rod}.
\end{eqnarray}

Combining these numbers with the relevant data from Table 4,
I find that, {\small \bf regardless of the level of~\mbox{activity}
or the lack of it}, the momentum effect driven by~the~sublimation of
water ice should vary as $r^{-3.0}$ and the production rate as
$r^{-3.2}$.  As explicitly remarked by Micheli et al.\ (2018), {\small \bf
this rate of variation is ``strongly disfavored by the data''}.

Although the present scenario is different from Micheli et al.'s
(2018) model, the predicted total {\it mass\/} production rates at
1.4~AU from the Sun agree to~within~20~percent of each other:\ I find
3.0~kg~s$^{-1}$ (for the disk case), whereas Micheli et al.\ get
3.6~kg~s$^{-1}$ when the contributions from water ice and carbon
monoxide are added~up.  Both models are comparable in their flexibility
to accommodate the observational constraints in terms of the upper limits
of the bulk density and the fraction of the surface that is active.  On the
other hand, because of the ``ray tracing'' of migrating activity over the
surface, the rigidity of Seligman et al.'s (2019) model does not allow
the option of relegating outgassing to a tiny fraction of the surface and
their model, requiring a water production rate as high as 9~kg~s$^{-1}$
(\mbox{3\,$\times$\,10$^{26}$\,molecules s$^{-1}$}), is out of line with
the production-rate constraints, as already noted briefly in Section~3.
These authors do not address the problem with the sublimation law and there
are other problems with their model, the two most worrisome being:\ (i)~the
derived dimensions of the object's model violate,
contrary to their claim, the constraints set by the nondetection of
`Oumuamua~by the Spitzer Space Telescope (Trilling et al.\
2018);\footnote{The relevant upper limit provided by Trilling et al.\ (2018)
for `Oumuamua's {\it semimajor\/} axis, based on the object's nondetection
by the Spitzer Space Telescope, is 171~meters, compared with 260~meters derived
by Seligman et al.\ (2019).}~and (ii)~the carbon-to-oxygen ratio obtained
from their water production rate and Trilling
et al.'s upper limits on the production rates of carbon dioxide and
carbon monoxide reaches an extremely low upper limit of \mbox{C/O\,$< 0.0003$},
the authors' value of 0.003 resulting from their two-orders-of-magnitude
error in quoting Trilling et al.'s constraint on the production of
CO.\footnote{Incredibly, both the disparity noted in footnote 9 and
the~misquoting of Trilling et al.'s (2018) upper limit of $Q_{\rm CO}$ passed
peer review of Seligman et al.'s (2019) paper uncorrected.}  Among the
18 comets with the H$_2$O, CO$_2$, and CO production rates tabulated
by Ootsubo et al.\ (2012), the lowest C/O ratio is $\sim$0.04, and with a few
exceptions CO$_2$ is more abundant than CO, in line with
Trilling et al.'s numbers; for the only Oort Cloud comet on Ootsubo et
al.'s list, C/2007~N3 (Lulin), the C/O ratio is between 0.08 and 0.10
and CO$_2$/CO ratio is $>$5.  Comets with extremely low ratios of C/O,
of $<$0.001, if they exist at all, must be exceptionally
rare:\ Ootsubo et al.\ do not have any on their list, Reach et al.\ (2013)
do very few, all highly uncertain and therefore dubious.

The problem of the nongravitational law actually illustrates a high
quality of both the astrometric observations employed by Micheli et
al.\ (2018) and their orbital analysis.  I note with surprise a recent
paper by Katz (2019), who questions Micheli et al.'s results.  His
objections are of two types, physical and methodical.  While I understand
his uneasiness about an implied dust-to-gas ratio of $<$0.0005, I see
{\small \bf no substantiation} for a conclusion that {\small \bf
`Oumuamua's orbital motion was affected by no (or nearly no) force other
than the Sun's gravity}.  While my impression is that the presence of a
nongravitational acceleration was unequivocally and convincingly
demonstrated by Micheli et al., Katz's doubts should be confronted
with independent evidence.  Since the set of more than 200
astrometric positions was supplied by nearly 30~groups of
observers, it is only the method of orbital analysis, which, if
overreaching, could make a difference.  In the following, Micheli
et al.'s findings are compared with the orbit determinations by other
authors.

Nakano (2018) used his own computer code to investigate `Oumuamua's
orbital motion, tabulating individual residuals (listed unfortunately
only to the nearest 0$^{\prime\prime\!}$.1) from two orbital solutions:\
a gravitational one and Marsden et al.'s (1973) Style~II nongravitational
one, solving for the radial and transverse components of the acceleration.
The residuals are quite small,~with~the mean residual of
$\pm$0$^{\prime\prime\!}$.42 for the gravitational solution and
$\pm$0$^{\prime\prime\!}$.33 for the nongravitational solution.  A
greater difference between both solutions is in the {\it systematic
trends\/} displayed by the residuals, which are most consistent and
therefore best seen among the positions derived from images taken with
the Hubble Space Telescope (HST) on 2017 November~\mbox{21--22} (15~data
points), December~12 (5~data points), and 2018 January~2 (10~data points).
{\vspace{-0.03cm}}Measured with a precision to 0$^{\prime\prime\!}$.001 and
having estimated errors of not more than 0$^{\prime\prime\!}$.04 according to
Micheli et al.\ (2018), the HST positions leave, from Nakano's gravitational
solution, steady residuals of +0$^{\prime\prime\!}$.1 in right ascension
and +0$^{\prime\prime\!}$.1 to +0$^{\prime\prime\!}$.2 in declination on
November~\mbox{21--22}; of +0$^{\prime\prime\!}$.4 to +0$^{\prime\prime\!}$.5
in right ascension~and +0$^{\prime\prime\!}$.1 to +0$^{\prime\prime\!}$.2 in
{\vspace{-0.01cm}}declination on December~12; and of $-$0$^{\prime\prime\!}$.1 to
$-$0$^{\prime\prime\!}$.2 in right ascension and $-$0$^{\prime\prime\!}$.2
in declination on January~2. On the other hand, the same HST positions leave,
from Nakano's nongravitational solution, equally consistently the residuals
of 0$^{\prime\prime\!}$.0 in both right ascension and declination on
November~\mbox{21--22}; of +0$^{\prime\prime\!}$.1 in right ascension and
0$^{\prime\prime\!}$.0 in declination on December~12; and of 0$^{\prime\prime\!}$.0
in right ascension and $-$0$^{\prime\prime\!}$.1 in declination on January~2.

To summarize:\ the unsatisfactory residuals that~are~left by Nakano's {\it
gravitational\/} solution from the HST positions absolutely demand the
presence of a nongravitational acceleration, fully confirming the conclusion
by Micheli et al.\ and not justifying Katz's doubts.  Whereas the residuals
left by Nakano's {\it nongravitational\/} solution are much better, they
are not systematic-trend free, suggesting, again in agreement with Micheli
et al.'s conclusion, that the Style~II nongravitational law does not provide
an optimum solution either, thereby {\small \bf dismissing the sublimation of water
ice as the trigger capable of precipitating the orbital perturbation}, which remains
unaccounted for in the outgassing model.

On the other hand, as a fair approximation,~Nakano's nongravitational
solution is expected to provide~a~meaningful estimate for the nongravitational
parameters.  For the parameter of the acceleration's radial component he found
\mbox{$A_1 = (+$5.03\,$\pm$\,0.11)$\, \times\:\! 10^{-4}$\,cm s$^{-2}$},
{\vspace{-0.03cm}} as opposed to Micheli et al.'s
\mbox{(+4.92\,$\pm$\,0.16)$\,\times\:\! 10^{-4}$\,cm s$^{-2}$},
within 1$\sigma$ of each other, while the transverse component's parameter
equaled \mbox{$A_2 = (-$0.055\,$\pm$\,0.042$)\,\times 10^{-4}$\,cm
s$^{-2}$}, suggesting that it was essentially zero and should not have been
solved for, once again in agreement with Micheli et al.'s findings.  Thus,
Nakano's results offer no evidence to compromise Micheli et al.'s results.

The task to compute an orbit of `Oumuamua was also undertaken by B.\ J.\ Gray
({\it Project Pluto\/}), whose elements and residuals (given to at least
{\vspace{-0.04cm}}0$^{\prime\prime\!}$.001) from the same 200+ observations are
online.\footnote{See {\tt https://projectpluto.com/temp/2017u1\_a.htm}.}
Gray was more strict with the observations than Nakano, rejecting 14
of those showing the residuals that exceeded 1$^{\prime\prime}$ (but not all
of them); his solution has a mean residual of $\pm$0$^{\prime\prime
\!}$.278.  The nongravitational effect is expressed in terms of an
``area-to-mass ratio'', which assumes an $r^{-2}$ law.  The
obtained value of 10.446\,$\pm$\,0.346~cm$^2$~g$^{-1}$ is equivalent to a
radial acceleration of (+4.74\,$\pm$\,0.16)\,$\times$\,10$^{-4}$\,cm
s$^{-2}$ at 1~AU from the Sun, within approximately 1$\sigma$ of Micheli et al.'s
(2018) result.  The residuals from the HST observations are of interest
for comparison with Nakano's; they range from $-$0$^{\prime\prime\!}$.001
to $-$0.$^{\prime\prime\!}$.012 in right ascension and from +0$^{\prime
\prime\!}$.005 to +0$^{\prime\prime\!}$.019 in declination on  November
21--22; from $-$0$^{\prime\prime\!}$.009 to +0$^{\prime\prime\!}$.039
in right ascension and from $-$0$^{\prime\prime\!}$.037 to +0$^{\prime
\prime\!}$.003 in declination on December~12; and from $-$0$^{\prime
\prime\!}$.018 to +0$^{\prime\prime\!}$.018 in right ascension and from
$-$0$^{\prime\prime\!}$.019 to +0$^{\prime\prime\!}$.007 in declination on
January~2.  Thus, the systematic trends are now, in the worst case, at a
level of 0$^{\prime\prime\!}$.01 to 0$^{\prime\prime\!}$.02, or about one
order of magnitude smaller than Nakano's (2018).  Note that if rounded off
to one decimal only, all HST residuals would be zero(!) and that Gray's
solution uses Micheli et al.'s preferred nongravitational law.

Another orbit determination effort is by Micheli's first co-author D.~Farnocchia.
The results are regarded as independent because the nongravitational solution,
presented on the website of the {\it JPL Small-Body~Database Browser},\footnote{See
Orbit 16 on {\tt https://ssd.jpl.nasa.gov/sbdb.cgi}.} employs a new sublimation law,
\begin{equation}
g(r) = 0.04084 \left( \frac{r}{5} \right)^{\!-2} \! \left[
  {\mbox{\raisebox{0ex}[3ex][3ex]{1}}} \!+\!
  \left( \frac{r}{5} \right)^3 \right]^{\!-2.6} \!\!,
\end{equation}
which should better accommodate the observations than the Style~II law.
All three components of the nongravitational acceleration were included in the
solution, with the parameters
\begin{eqnarray}
A_1 & = & (+5.59 \pm 0.72) \times 10^{-4}\,{\rm cm\;s}^{-2}, \nonumber \\
A_2 & = & (+0.29 \pm 0.49) \times 10^{-4}\,{\rm cm\;s}^{-2}, \nonumber \\
A_3 & = & (+0.32 \pm 0.45) \times 10^{-4}\,{\rm cm\;s}^{-2}.
\end{eqnarray}
Obviously, $A_2$ and $A_3$ are meaningless.  The mean residual, $\pm$0$^{\prime
\prime\!}$.436, is fairly large, possibly because \mbox{Farnocchia} included 15~more
observations in the solution than Gray, some of them presumably leaving larger
residuals.~However, since the individual residuals are unavailable on the
website, this cannot be stated with certainty.  The larger error of $A_1$ could
be an effect of the correlations between the nongravittaional parameters
and the same might be true about the value of $A_1$.  Nonetheless,
even so the deviation of $A_1$ from 4.92\,$\times$\,10$^{-4}$~cm~s$^{-2}$ is less
than 15~percent.  I conclude that {\small \bf Micheli et al.'s results are
confirmed by all three presented orbital analyses; comparison of the HST residuals
from Nakano's and Gray's solutions shows that the inverse square law is superior
to the Style~II law, in line with independent evidence implying that the sublimation
of water ice~is not triggering the nongravitational acceleration}.

In the context of `Oumuamua's extremely low bulk density inferred from the
considerations in the previous sections, it is noted that Kataoka et al.\ (2013)
provide a pathway for long-term growth of a fluffy aggrerate of submicron-sized dust
grains into a planetesimal in a protoplanetary disk (see also Kataoka 2017).
{\vspace{-0.03cm}}According to these authors, the initial assembly of grains, each
10$^{-15}$\,g in mass and of material density, reaches eventually a terminal mass
of 10$^{18}$\,g and a diameter of 10~km.  The earliest stage of the evolution is
dominated by low-velocity collisional coagulation, in which the bulk density of the
dust aggregate of steadily augmenting dimensions keeps dropping as an inverse square
root of mass (i.e., with a mass fractal dimension of 2).  Growth of the aggregate
gradually invokes collisional compression, whose effect however is not enough to
reverse the trend toward increasing porosity until the bulk density drops to as low
as 0.00002~g~cm$^{-3}$, at which time the aggregate is $\sim$10$^{-5}$\,g in mass
and nearly 1~cm across.  The aggregate's slow compaction, leading to an increase
in the bulk density by about one order of magnitude by the time the mass is some
100~g and diameter 1~m, is largely a product of quasi-static compression by ambient
gas in the protoplanetary disk.  However, in the mass range from 10~kg to a million
tons (spanning eight orders of magnitude!) the bulk density is remarkably flat,
staying between 0.0002 and 0.0004~g~cm$^{-3}$ at all times until self-gravity sets
in, when the object is hundreds of meters across and getting rapidly denser from
that point on.

Assuming that `Oumuamua was the largest inactive fragment of an interstellar
comet that was an embryo planetesimal in a stage of evolution that had preceded~the
self-gravity compression stage, the former's mass should be a non-negligible
fraction of the latter's.  For the parent's assumed diameter of $\sim$120~m and bulk
density~of $\sim$0.0003~g~cm$^{-3}$ in this stage of evolution, its mass was equal
to a few{\vspace{-0.04cm}} times 10$^8$\,g in Kataoka's (2017) plot, so that the
mass of 10$^7$\,g, proposed for `Oumuamua's fluffy model (Sekanina 2019a), fits
the notion of the parent's ``skeleton'' residuum, amounting to several percent of
the mass upon arrival at the inner Solar System in early 2017, a plausible outcome
of the parent's putative disintegration near perihelion.

\section{Conclusions}
At the time of Micheli et al.'s (2018) paper on the nongravitational effects
in `Oumuamua's orbital motion, there were only two warning signs that made
activity highly questionable as the acceleration's trigger:\ one was the
object's star-like appearance, which indicated an extremely low level
of dust production (Meech et al.\ 2017); the other was the strict constraint
on~the~production~of CN, C$_2$, and C$_3$ (Ye et al.\ 2017).~For~some
reason,~neither argument turned out to be strong enough~to~\mbox{deter}
the~\mbox{proponents}.  A third warning sign,~the~\mbox{sublimation} law,
was brought up by Micheli et al.\
themselves,~but~it related only~to~the production of water and it~seems~that
its signifiance was downplayed.  The most damaging finding to the outgassing
hypothesis was the very tight limit of the abundances of CO$_2$ and CO derived
 by Trilling et al.\ (2018), not available until months~later.~With~both
carbon-bearing volatiles eliminated as the prime mover of activity,
water ice remained the only hope of the outgassing hypothesis.

This paper elaborates on the grave difficulties one confronts in an attempt
to identify the sublimation of water ice with `Oumuamua's presumed activity and
with the trigger of the detected nongravitational acceleration:\

(a)~the variation with heliocentric distance is incompatible with the
law that fits the orbital data;

(b)~the acceleration exceeds the outgassing-driven accelerations
of practically all comets, while the star-like appearance of `Oumuamua
is in sharp contrast with~the considerable diffuseness of C/1998~P1 and other
comets with sizable (yet presumably outgassing-induced)
nongravitational effects; and

(c)~the absence of detectable activity is shown
in Sections 5--6 to imply that the sublimation of water ice proceeded, if it did
at all, from only a very small fraction of the surface and that the bulk density
would have to be extremely low, below 0.001~g~cm$^{-3}$; the porosity and
morphology would then readily fit the scenario in which `Oumuamua was described as
an inactive fragment of a dwarf interstellar comet that had disintegrated near
perihelion weeks before discovery (Sekanina 2019a).  This hypothesis is also
consistent with the parent comet having been an embryo planetesimal in the
coagulation growth pathway of fluffy dust aggregates constructed by Kataoka et al.\
(2013) and Kataoka (2017).   The nongravitational acceleration of `Oumuamua was
then driven by solar radiation pressure, in which case there indeed should be
no need for even a trace of activity; not to mention that the difficulties
under (a) and~(b)~disappear by default.

While I see no avenue for outgassing in general,~and the sublimation of
water ice in particular, to perform as a trigger for `Oumuamua's
nongravitational~acceleration, it should be acknowledged that it is the
high-quality imaging, image processing, astrometric work, as well as
the superb orbital analysis available that made~it~possible to constrain the model
to this extent. \\

This research was carried out at the Jet Propulsion Laboratory, California
Institute of Technology, under contract with the National Aeronautics and
Space~Administration.\\[1.84cm]
%
%
\pagebreak
\begin{center}
{\footnotesize REFERENCES}
\end{center}
\vspace{-0.3cm}
\begin{description}
{\footnotesize
\item[\hspace{-0.3cm}]
A'Hearn, M.\ F., Millis, R.\ L., Schleicher, D.\ G., et al.\ 1995, Icarus,{\linebreak}
 {\hspace*{-0.6cm}}118, 223
\\[-0.57cm]
%
%
\item[\hspace{-0.3cm}]
Bair, A.\ N., Schleicher, D.\ G., \& Knight, M.\ M.\ 2018, AJ, 156, 159{\hspace{0.2cm}} 
\\[-0.57cm]
%
%
%
\item[\hspace{-0.3cm}]
Bodewitz, D., Villanueva, G.\ L., Mumma, M.\ J., et al.\ 2011, AJ,{\linebreak}
 {\hspace*{-0.6cm}}141, 12 
\\[-0.57cm]
%
%
\item[\hspace{-0.3cm}]
Do, A., Tucker, M.\ A., \& Tonry, J.\ 2018, ApJ, 855, L10 
\\[-0.57cm]
\item[\hspace{-0.3cm}]
Drahus, M., Guzik, P., Waniak, W., et al.\ 2018, Nature Astron., 2,{\linebreak}
 {\hspace*{-0.6cm}}407
\\[-0.57cm]
\item[\hspace{-0.3cm}]
Farnham, T.\ L., Schleicher, D.\ G., Woodney, L.\ M., et al.\ 2001,{\linebreak}
 {\hspace*{-0.6cm}}Science, 292, 1348
\\[-0.57cm]
%
%
\item[\hspace{-0.3cm}]
Hale, A.\ 2000, Int.\ Comet Quart., 22, 44
\\[-0.57cm] 
%
%
\item[\hspace{-0.3cm}]
Kataoka,\,A.\,2017, in Formation, Evolution, and Dynamics of Young{\linebreak}
 {\hspace*{-0.6cm}}Solar Systems, ed.\ M.\ Pessah \& O.\ Gressel, Astrophysics
 and{\linebreak}
 {\hspace*{-0.6cm}}Space Science Library, 445 (Cham, Switzerland:\ Springer), 143
\\[-0.57cm]
\item[\hspace{-0.3cm}]
Kataoka, A., Tanaka, H., Okuzumi, S., \& Wada, K.\ 2013, A\&A,{\linebreak}
 {\hspace*{-0.6cm}}557, L4
\\[-0.57cm]
\item[\hspace{-0.3cm}]
Katz, J.\ I.\ 2019, Astrophys.\ Space Sci., 364, 51
\\[-0.57cm]
%
%
\item[\hspace{-0.3cm}]
Knight, M.\ M., \& Schleicher, D.\ G.\ 2015, AJ, 149, 19 
\\[-0.57cm]
%
%
\item[\hspace{-0.3cm}]
Lupu, R.\ E., Feldman, P.\ D., \& Weaver, H.\ A.\ 2007, ApJ, 670, 1473
\\[-0.57cm]
\item[\hspace{-0.3cm}]
Marsden, B.\ G., \& Williams, G.\ V.\ 2008, Catalogue of Cometary{\linebreak}
 {\hspace*{-0.6cm}}Orbits 2008, 17th ed.\ (Cambridge, MA:\ IAU Minor Planet
 Cen-{\linebreak}
 {\hspace*{-0.6cm}}ter/Central Bureau for Astronomical Telegrams), 195pp
\\[-0.57cm]
\item[\hspace{-0.3cm}]
Marsden, B.\ G., Sekanina, Z., \& Yeomans, D.\ K.\ 1973, AJ, 78,~211
\\[-0.57cm]
\item[\hspace{-0.3cm}]
Meech, K.\ J., Weryk, R., Micheli, M., et al.\ 2017, Nature, 552, 378
\\[-0.57cm]
\item[\hspace{-0.3cm}]
Micheli, M., Farnocchia, D., Meech, K.\ J., et al.\ 2018, Nature, 559,{\linebreak}
 {\hspace*{-0.6cm}}223
\\[-0.54cm]
\item[\hspace{-0.3cm}]
Mumma,\,M.\,J., Dello Russo,\,N., DiSanti,\,M.\,A., et al.\,2001,~Science,{\linebreak}
 {\hspace*{-0.6cm}}292, 1334
\\[-0.57cm]
%
%
\item[\hspace{-0.3cm}]
Nakano, S. 2018, NK 3691
\\[-0.57cm]
\item[\hspace{-0.3cm}]
Ootsubo, T., Usui, F., Kawakita, H., et al.\ 2010, ApJ, 717, L66
\\[-0.57cm]
\item[\hspace{-0.3cm}]
Ootsubo, T., Kawakita, H., Hamada, S., et al.\ 2012, ApJ, 752, 15 
\\[-0.57cm]
\item[\hspace{-0.3cm}]
Park, R.\ S., Pisano, D.\ J., Lazio, T.\ J.\ W., et al.\ 2018, AJ, 155, 185
\\[-0.57cm]
\item[\hspace{-0.3cm}]
Probstein, R.\ F.\ 1969, in Problems of Hydrodynamics and Con-{\linebreak}
 {\hspace*{-0.6cm}}tinuum Mechanics, ed.\ F.\ Bisshopp \& L.\ I.\ Sedov
 Philadelphia:{\linebreak}
 {\hspace*{-0.6cm}}Soc.\ Ind.\ Appl.\ Math.), 568
\\[-0.57cm]
%
%
%
\item[\hspace{-0.3cm}]
Reach, W.\ T., Kelley, M.\ S., \& Vaubaillon, J.\ 2013, Icarus, 226, 777
\\[-0.57cm]
%
%
\item[\hspace{-0.3cm}]
Sekanina, Z.\ 1982, in Comets, ed.\ L.\ L.\ Wilkening (Tucson, AZ:{\linebreak}
 {\hspace*{-0.6cm}}University of Arizona), 251
\\[-0.57cm]
\item[\hspace{-0.3cm}]
Sekanina, Z.\ 1993, AJ, 105, 702
\\[-0.57cm]
%
%
%
\item[\hspace{-0.3cm}]
Sekanina, Z.\ 2019a, eprint arXiv:1901.08704
\\[-0.57cm]
\item[\hspace{-0.3cm}]
Sekanina, Z.\ 2019b, eprint arXiv:1903.06300
\\[-0.57cm]
%
%
\item[\hspace{-0.3cm}]
Sekanina, Z., \& Kracht, R.\ 2015, ApJ, 801, 135
\\[-0.57cm]
%
%
\item[\hspace{-0.3cm}]
Seligman, D., Laughlin, G., \& Batygin, K.\ 2019, ApJ, 876, L26
\\[-0.57cm]
%
%
\item[\hspace{-0.3cm}]
Trilling, D.\ E., Mommert, M., Hora, J.\ L., et al.\ 2018, AJ, 156,~261
\\[-0.57cm]
%
%
%
%
\item[\hspace{-0.3cm}]
Yabushita, S.\ 1996, MNRAS, 283, 347
\\[-0.65cm]
\item[\hspace{-0.3cm}]
Ye, Q.-Z., Zhang, Q., Kelley, M.\ S.\ P., \& Brown, P.\ G.\ 2017, ApJ,\\[-0.08cm]
 {\hspace*{-0.6cm}}851, L5}
\vspace{1.79cm}
\end{description}
\end{document}